\documentstyle[12pt,epsfig]{article}
\newlength{\dinwidth}                       
\newlength{\dinmargin}                      
\def\lsim{\mathrel{\rlap{\lower4pt\hbox{\hskip1pt$\sim$}}
    \raise1pt\hbox{$<$}}}                % less than or approx. symbol
\def\gsim{\mathrel{\rlap{\lower4pt\hbox{\hskip1pt$\sim$}}
    \raise1pt\hbox{$>$}}}                % greater than or approx. symbol
\def\Journal#1#2#3#4{{#1} {\bf #2} (#4) #3}
\def\zz{\vspace*{-1.3mm}}

\def\NP{{ Nucl.\ Phys.\ }}
\def\PL{{ Phys.\ Lett.\ }}

\def\PR{{ Phys.\ Rev.\ }}
\def\ZP{{ Z. Phys.\ }}
\newlength{\gap}
\begin{document}

\begin{center} 
{\begin{flushright}
SINP-96-71\ \ \ \  \ \ \ hep-th/9608295
\end{flushright}
\vspace*{3mm}
 \Large \bf
 Higgs production and other Two-Photon Processes\\
 in eA Collisions at HERA energies}  

\vspace*{4mm}
\begin{large}
B.B.~Levtchenko
\end{large}

Institute of  Nuclear Physics, Moscow State University, 119899 Moscow, Russia\\     
\end{center}

\vspace*{-0mm}
\begin{quotation}
\noindent
{\bf Abstract:}
The planed HERA upgrade will increase the integrated luminosity by two orders of magnitude.
A new program of electron-nucleus collisions at HERA is  under discussion now.
In  this connection we present prospects for the study of a light neutral Higgs boson,
 SUSY-particle, hadrons and leptons production
in the two-photon coherent $eA$ collisions at HERA energies.
\end{quotation}

\section{Introduction}

\vspace*{-1mm}
Particle production in the two-photon processes in
 lepton-lepton, hadron-hadron and nucleus-nucleus collisions
 has been  already discussed in the literature \cite{BGMS}-\cite{papag1}.
One of the goals of the present analysis is  to study the discovery potential
of electron-nucleus collisions at  HERA  in the case of coherent electromagnetic 
 particle production.  

Requirement of coherency implies that the collision is quasi-elastic,  the  impact
parameter $b$  is larger than the nuclear radius and the nuclear charge acts as a whole.
    For such collisions the hadronic background
from  $\gamma^{*} q$ and $\gamma^{*} g$  (direct and resolved)
 subprocesses will be absent  offering a better 
experimental environment for studying new particle production.

Due to the large charge $Z$ of heavy ion and the strong contraction of the
electromagnetic fields in a high-energy collision we expect in such
 an electron-nucleus interaction a large  particle
 production  rate  by the two-photon fusion mechanism. It turns out  that the cross sections, which is scale
with $Z^2$, are comparable or even large than the total  cross sections expected in
$e^{+}e^{-}$ ,  ep and  pp collisions of the same incident energy per particle.

An important characteristic of $eA$ collider is it discovery range:
how heavy  produced particles  can be and how large will be production
cross sections for them.  In the  equivalent photon approximation
the invariant mass squared of the two-photon system, denoted by $W^2$, is
fixed by the photon energies $\omega_{e}$ and $\omega _{A}$. 
As it well known 
\cite{ BB}, \cite{papag1} that
the time duration of the electromagnetic pulse, produced by a charge moving with a relativistic
$\gamma$  factor, corresponds to field frequencies of $\omega \leq \gamma/b$. 
 In the case of a heavy-ion beam  the largest photon energy is
 $\omega_{A} \simeq \gamma_{A}/R_{A}$ although for photons emerged from an
electron beam the photon energy can be close to the electron energy 
$\omega_{e} \leq  E_{e}$.  Thus the discovery range of  the  $eA$ HERA collider for
coherent processes extends to masses:

\vspace*{-3mm}
\begin{equation}
\label{DR}
 W \approx  \left(4\omega_{e}\omega_{A}\right )^{1/2}\, \leq \left(\frac{4\,E_{e}E_{p}}{m_{p}R_{A}}\right )^{1/2}\,. 
\end{equation}
This means the invariant mass values accessible at HERA energies are  up
to  86 $GeV$,   71$GeV$ and 54 $GeV$
  for  $eC$, $eCa$ and $ePb$ collisions, respectively.

The paper is organized as follows:
In the next section we compare the discovery ranges   
of  $eA$ (HERA),   $e^{+}e^{-}$ (LEP2)  and $ep$ (HERA)  colliders
using  the two-photon luminosity function . 
In Sect.3, we evaluate cross sections for coherent production of lepton and slepton pairs 
($e^{+}e^{-}$, $\mu^{+}\mu^{-}$, $\tau^{+}\tau^{-}$, $\tilde{l^{+}}\tilde{l^{-}}$), charged pion pair, 
 pseudoscalar
($J^{C}=0^{+}$)  resonances and the MSSM light  Higgs boson.
Conclusions are given in the last section.

\vspace*{-1mm}
\section{ Two-photon luminosity function}

The equivalent photon approximation   is allow  to express  the cross section  for the
two-photon production processes in $eA$ collisions as follows \cite{BGMS}, \cite{Kessler1}

\begin {equation}
\label{Xs}
d\sigma_{eA}\  = \ \sigma_{\gamma\gamma} dn_{e}dn_{A}
\end{equation}
where $dn_{i}$ is the number of equivalent photons, or a distribution in photon frequencies.
It is convenient to define \cite{Field2}  a dimensionless quantity  $L_{\gamma\gamma}$  called 
the two-photon luminosity (TPL)
\begin {equation}
\label{TPL-G}
 dL_{\gamma\gamma} = \int dn_{e}(x_{e}) dn_{A}(x_{A})\delta(z^{2}-x_{e}x_{A}), 
\end{equation}
where $z=W/\sqrt{s_{eA}}$ and $x_{i}=2\omega_{i}/\sqrt{s_{eA}}$ . Function $L_{\gamma\gamma}$ depends  on the
type of colliding particles and Lorentz invariant quantities  and is very useful for comparing 
the relative efficiency of the different colliders\footnote{
 To get the production rate Eq. (\ref{Xs}) have to be multiplied
by the machine luminosity $\mathcal{L}$$_{eA}$ [$cm^{-2}sec^{-1}$].  For ion beams 
 $\mathcal{L}$$_{eA}\sim  A^{-1} \mathcal{L}$$_{ep}  $.}.

Let to  calculate the  TPL function in the center of mass of colliding an  electron and heavy-ion.
If we integrate eq. (\ref{TPL-G}) over the energy range of one of the photons
$ t^{*}\leq\,x_{e}\,\leq\,1$  with the photon spectra 

\begin{equation}
dn_{e}\, =\, \frac{\alpha}{2\pi}\, [1\,+\,(1\,-\,x_{e})^{2}]\,\ln\frac{s}{4m^{2}_{e}}\,\frac{dx_{e}}{x_{e}} 
\end{equation}
 of an electron \cite{ Low}, \cite{BGMS} and

\begin{equation}
dn_{A}\, =\,\frac{2Z^{2}_{A}\alpha}{\pi}\, \ln\frac{\gamma_{A}}{\omega_{A}\,R_{A}}\,\frac{dx_{A}}{x_{A}} 
\end{equation}
of an ion  \cite{BB} we obtain the following expression for the differential two-photon function:
\begin{equation}
\label{TPL}
\frac{dL_{\gamma\gamma}}{dW}\, = \,\frac{2Z^{2}_{A}\alpha^{2}}{\pi}\,\frac{f(t^{*})}{W}\,ln\left(\frac{s_{eA}}{4m^{2}_{e}}\right ),
\end{equation}
where
$$f(t^{*})= (\ln t^{*}+\frac{\sqrt{3}}{2})^{2}+(t^{*}-4)^{2}-3  $$
and
$t^{*}\,=\,\frac{R_{A}\sqrt{s_{eA}}}{2\gamma_A}\, z^{2}\,.$  Eq. (\ref{TPL}) is valid in the leading logarithm approximation.

Thus the differetial cross section  for the
production of a final state with the effective mass $W$ in the two photon $eA$ collision is a product

\vspace*{-1mm}
\begin {equation}
\label{xseA}
d\sigma_{eA}\  = \ \sigma_{\gamma\gamma}\cdot \frac{dL_{\gamma\gamma}}{dW}dW
\end{equation}
of  the cross section $\sigma_{\gamma\gamma}$ of the ${\gamma\gamma}\rightarrow\it f$ transition for
real non-polarized photon and the TPL function.

%\vspace*{-5mm}
\begin{figure}[h!]
\begin{tabular}{p{8cm} p{8cm}}
\mbox{\hspace{-2mm}
\epsfig{file=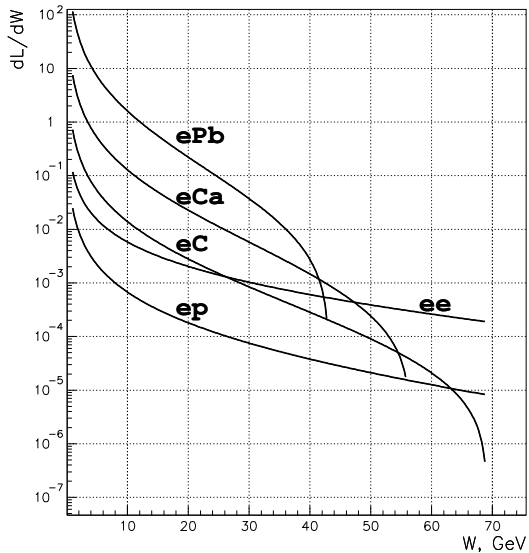 , width=8.0cm%
}}\vspace*{-6mm}
\caption{\it
Spectra of the  two-photon luminosity function for  
 $e^+e^-$ (LEP,  200 GeV), ep(HERA,  300 GeV),
 and $eA$ (HERA, 300A\,GeV) colliders. }&
\vspace*{-8.0cm}

In Fig.~1 shown the dependence of TPL on the two-photon invariant mass for different colliding beams.
The TPL function for $e^+e^-$ (LEP2 )  was calculated with using Eq.(6) of ref. \cite{Field2},
as for  the  $ep$ collision (HERA) the elastic part of the photon spectra
 from the proton was taken from \cite{ACK}.  Because of the used approximation the  range of
allowed $W$ values  in (\ref{TPL})   is narrow (see Fig.1) as compared to (1) and  defined by 
the condition  $f(t^{*})\geq0$.

The advantage of $eA$ collisions over $e^+e^-$ and $ep$ collisions lies in the $Z^2$ factor, which enchances the
photon flux up to a factor $10^4$. This allow, in the range of $W$ below 40 GeV, the study of 'new phenomena' 
at HERA with large statistics. Results presented here were obtained for untagged scattered electron and nucleus. 
The effect of the nuclear formfactors on the TPL function is not included.
\end{tabular} 
\end{figure}

%\vspace*{-3mm}
\section{Particle production processes}
In this section we present a few examples of particle production in $eA$ collisions 
through the two-photon mechanism. Processes with production of the 'old' particles 
(light hadrons and leptons) will 
create a background at search of 'new' particles (like Higgs bosons and sleptons).
Cross sections  presented here were calculated at $E_e=27\,$GeV and $E_A=\,E_p A$
with $E_p=820\,$GeV.

\vspace{2mm}
{\it Resonances and pions}
\vspace{2mm}

A region of  the  low mass in the two-photon system is the resonance region.
C-even pseudoscalar hadrons - $\pi^0,\ \eta,\ \eta^{'},\ \eta_c,\ \chi_{c0}$ - will be
produced copiously at $W<4\,$GeV.  Their production cross section $\sigma_{\gamma\gamma}$ is
well known \cite{BGMS}
\begin{equation}
\label{gg}
\sigma_{\gamma\gamma\rightarrow R}= 8\pi \frac{\Gamma^{\gamma\gamma}\Gamma}{(W^2-m^2_R)^2-\Gamma^2m^2_R}
\end{equation}
and expressed through the two-photon decay width $\Gamma^{\gamma\gamma}$ of the resonance R.
Now Eqs. (\ref{xseA}) and (\ref{TPL})  allows  easily to evaluate dependence of
the production rate on  $W$. 
Fig.~2a shows a resonance structure of the differential cross section (\ref{xseA}) in that region.
 
As another example  the cross section (\ref{xseA}) for pair production of charged pions  is plotted in Fig.~2b. 
The corresponding total $\gamma\gamma\rightarrow \pi^+\pi^-$ cross section in Born approximation
 (pointlike pions) \cite{BGMS}  is 
\begin{equation}
\label{pipi}
\sigma_{\gamma\gamma}\, =\, \frac{2\pi\alpha^2}{W^2}\,\left[\left(1+\frac{4m^2_{\pi}}{W^2} \right)\sqrt{1-\frac{4m^2_{\pi}}{W^2}}
\,-\,\frac{8m^2_{\pi}}{W^2}\left(1-\frac{2m^2_{\pi}}{W^2}\right)\ln\left(\frac{W}{2m_{\pi}}\,+\,\sqrt{\frac{W^2}{4m^2_{\pi}}-1}\right)\right].
\end{equation}
In \cite{BGMS} is also discussed in details what can be measured  in the above processes and how. 
We  note only,  that the study of the angular and energy correlations between
 pions allows, in principle, to extract 
components of the amplitude $M_{ab}\,(a,\,b\,=\,   +,\,-,\, 0)$ for the ${\gamma\gamma}\rightarrow\it f$ transition.

\vspace*{-2mm}
\begin{figure}[t]
\begin{tabular}{cc}
\epsfig{file=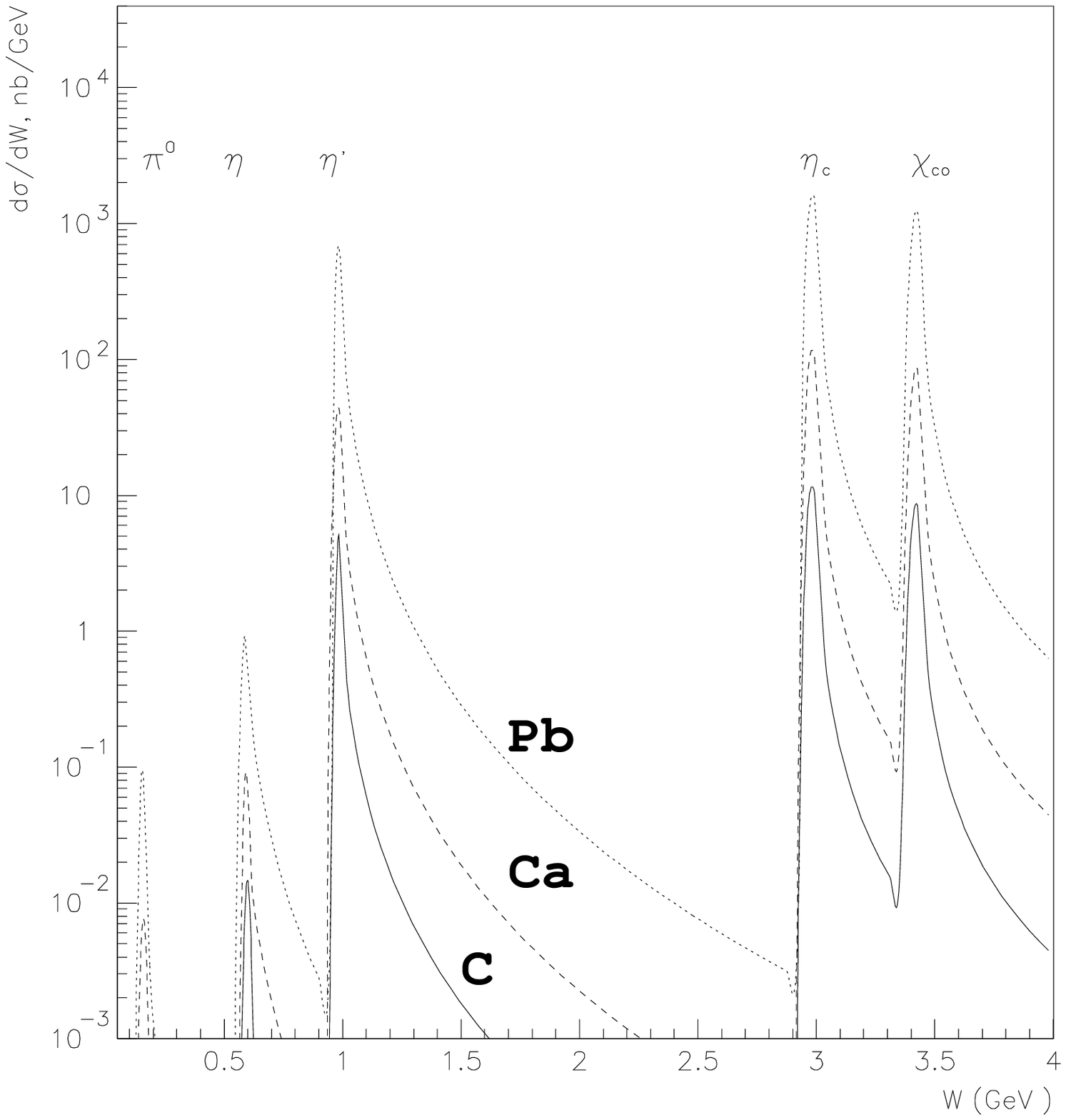,height=8.0cm}
\epsfysize=8cm
{\epsffile{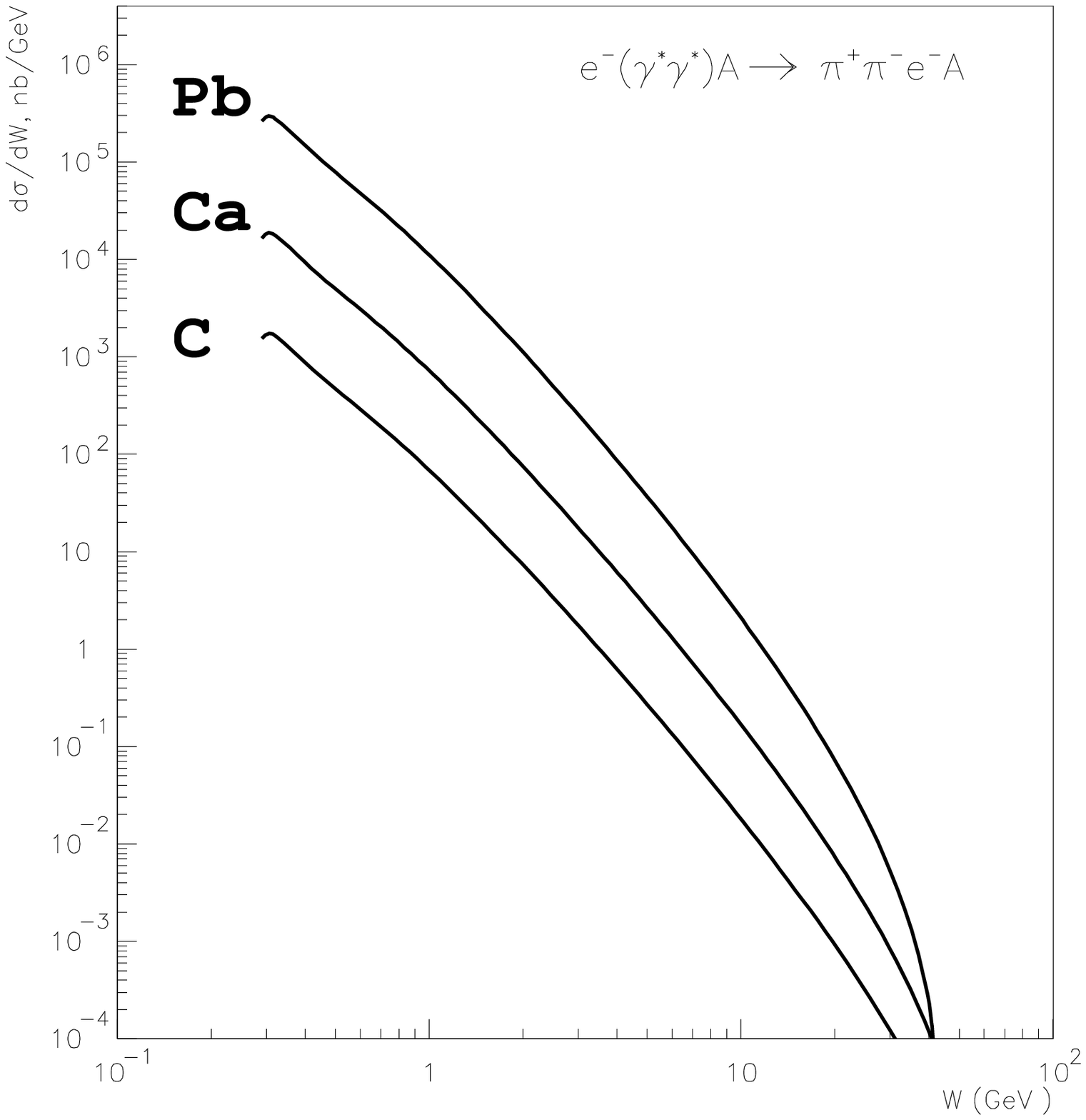}} 
\end{tabular}
\vspace*{-4mm}
\caption{\it
Differential cross section for production of C-event pseudoscalar hadrons
(a) and a $\pi^+\pi^-$ pair (b) as a function of $W$ and a nucleus size. }
\end{figure}
%=======================

\vspace{2mm}
{\it Lepton pairs}
\vspace{2mm}

The Bethe-Heitler  process is a source of lepton pairs  with extremely high yield (Fig.~3). 
Including the mass effects at the threshold, the cross section \cite{BGMS} for producing a pair of identical leptons is
\begin{equation}
\label{BH}
\sigma_{\gamma\gamma}\, =\, \frac{4\pi\alpha^2}{W^2}\, \left[2\left(1\,+\,\frac{4m^2_l}{W^2}\,-
\,\frac{8m^4_l}{W^4}\right)\ln\left(\frac{W}{2m_l}\,+\,\sqrt{\frac{W^2}{4m^2_l}-1}\right)
- \left(1+\frac{4m^2_l}{W^2} \right)\sqrt{1-\frac{4m^2_l}{W^2}}\right],
\end{equation}
where $m_l$ is the lepton mass.
Copious production of $\tau$-leptons provide an opportunity to study QCD in $\tau$ decay
\cite{tau}. However it will also to complicate the Higgs boson
search in  lepton decay modes because
the main background above the $\tau$  threshold is due to 
$\gamma^*\gamma^*\rightarrow\tau^+\tau^-$.
  How to suppress the Bethe-Heitler background is discussed in \cite{bussey} .

\vspace*{-0mm}
\begin{figure}[t]
\begin{tabular}{ccc}
\hspace {-1.5cm}
\epsfysize=7cm
{\epsffile{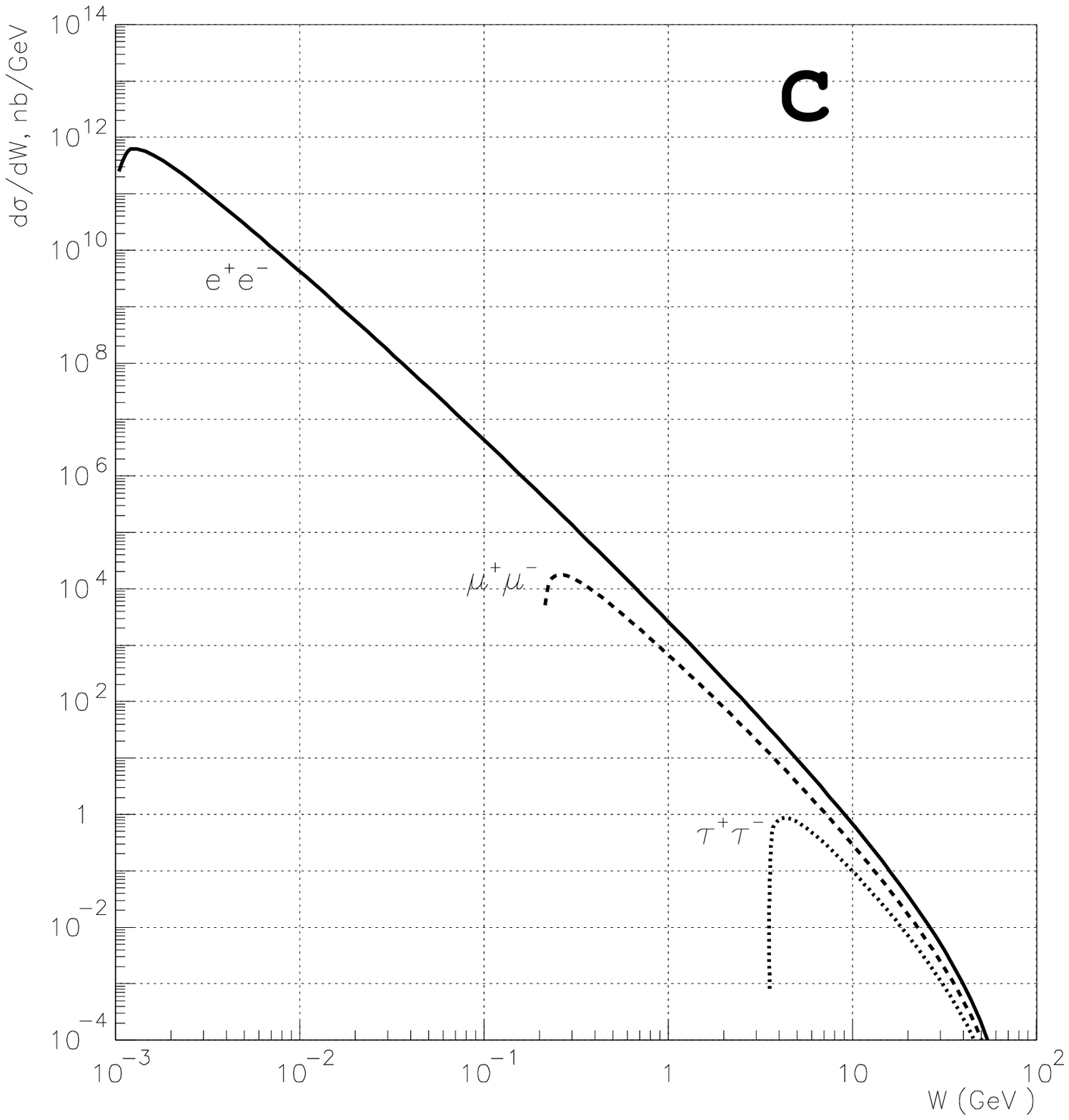}} &
\hspace {-1.0cm}
\epsfysize=7cm
{\epsffile{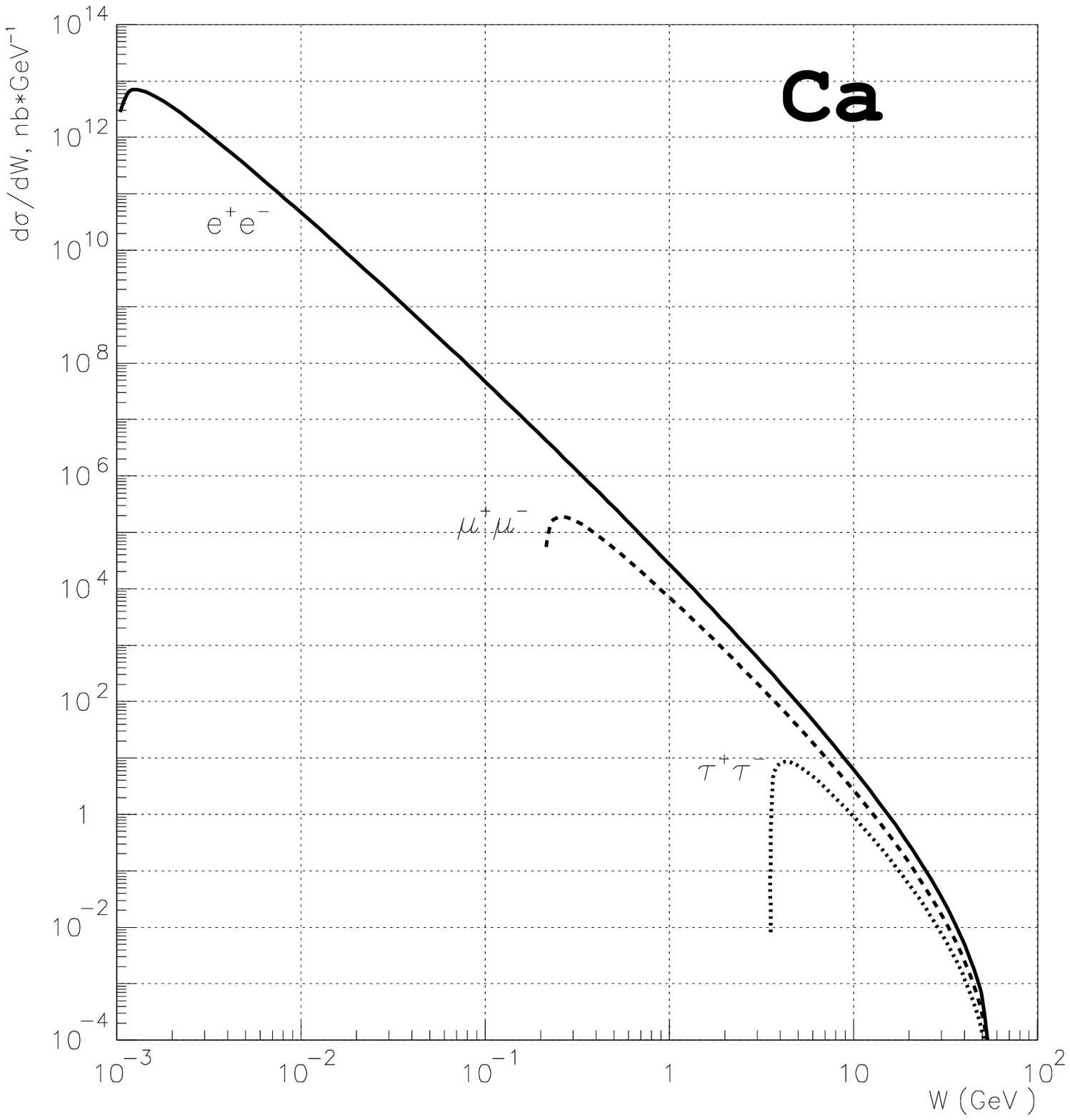}} &
\hspace {-1.cm}
\epsfysize=7cm
{\epsffile{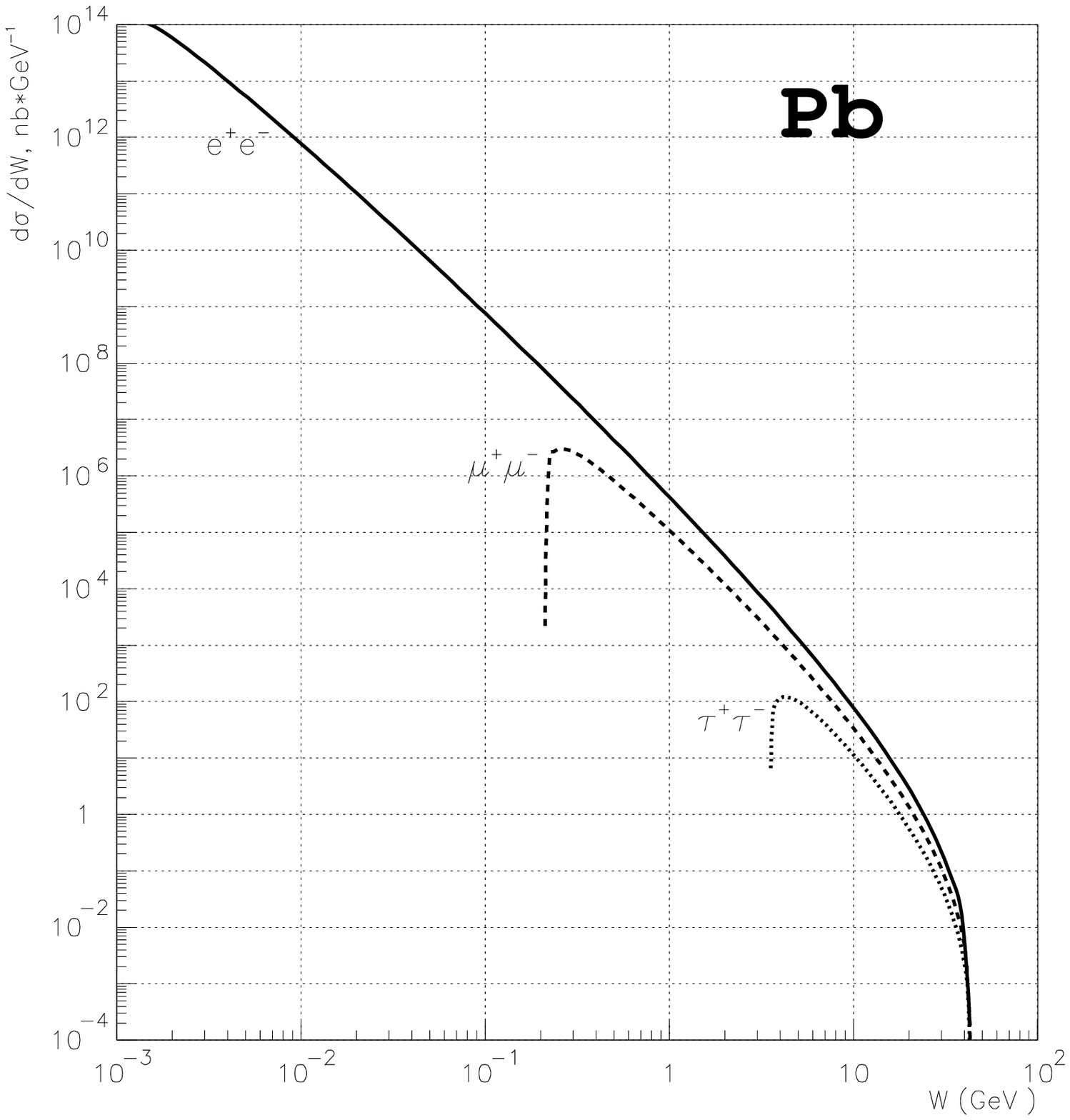}}
\end{tabular}
\vspace*{-4mm}
\caption{\it
Same as Fig.~2 for production of BH leptons in coherent eA collisions. }
\end{figure}
%=======================

\vspace{2mm}
{\it SUSY particles}
\vspace{2mm}

Non-strongly interacting sypersymmetric particles - sleptons and charginos-
can be detected at the $eA$ collider  discussed here if their masses lies in the
range $M_{sl}< 30$ GeV. Chargino couple to photons in the standard way of spin
1/2 fermions and sleptons are spin-zero particles.  The two-photon cross
sections for production a pair of them \cite{BKT} are coincide with eqs. (\ref{BH}) and (\ref{pipi}), respectively.
$\sigma(\gamma\gamma\rightarrow\tilde{l}^+\tilde{l}^-)$
 is not logarithmically enhanced thus the production rate for sleptons is
significantly lower than the chargino rate.
As an example ,  the total cross section for slepton production 
versus the mass of the produced particle  is shown in Fig.~4a.
  At an  integrated luminosity of 
$10^3/A\,$pb$^{-1}$  only sleptons with masses not heavier  29 GeV, 25 GeV and
20 GeV can be detected in $eC,$ $eCa$, $ePb$ collisions respectively. At
$\mathcal{L}$$_{eA}\sim 1/A$ pb$^{-1}$ the discovery range for sleptons shrinks to
$M_{\tilde{l}}<13$ GeV.

\vspace*{-0mm}
\begin{figure}[t!]
\begin{tabular}{cc}
%
%\hspace {-1.2cm}
\epsfysize=8cm
{\epsffile{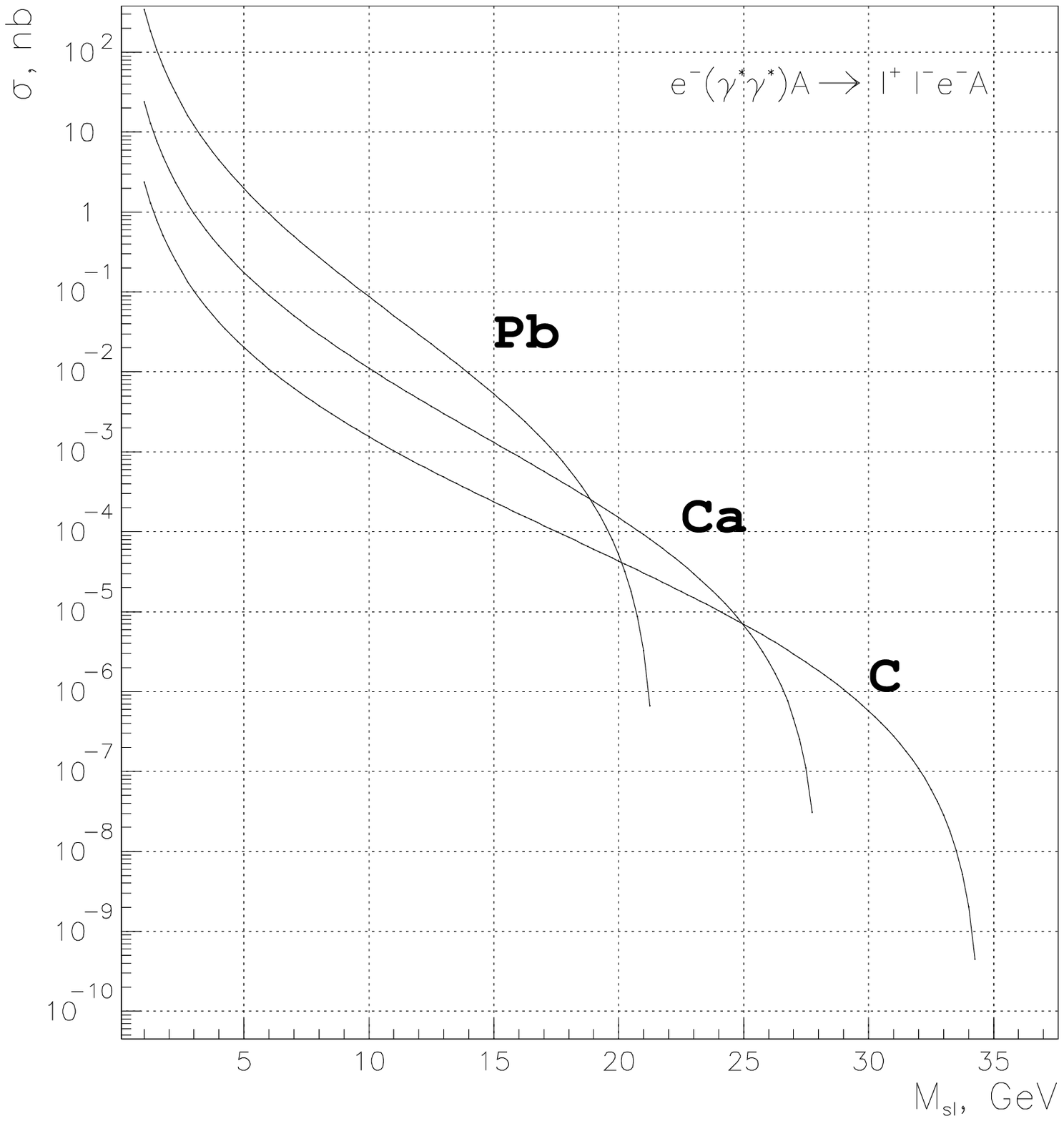}} &
%\hspace {-1.3cm}
\epsfysize=8cm
{\epsffile{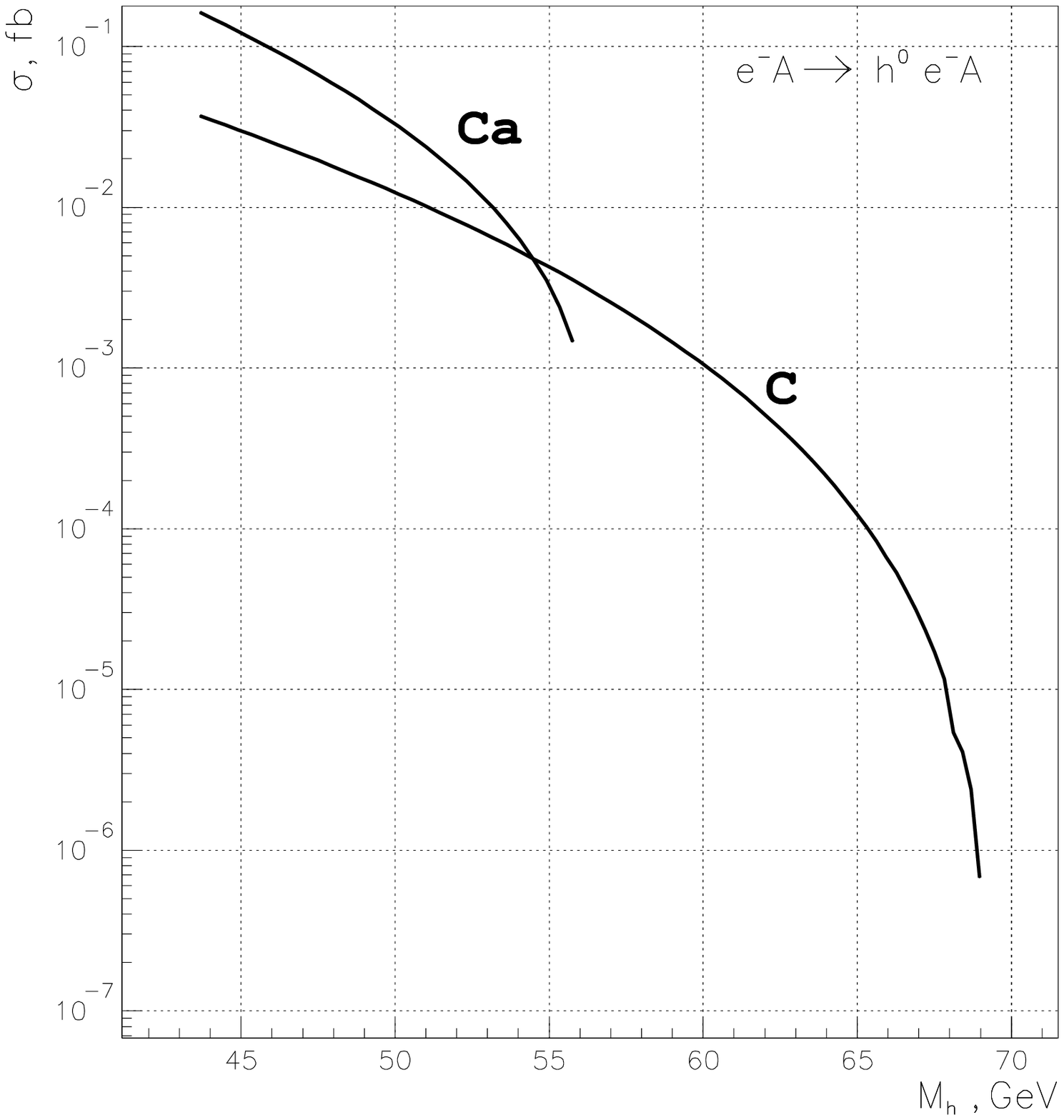}} 
\end{tabular}
\vspace*{-4mm}
\caption{\it
Cross sections for slepton (a) and  MSSM light Higgs  boson ($tan\,\beta =1.75$) (b)
production through two-photon fusion. The cross sections are
plotted versus the mass of the produced particle. }
\end{figure}
%=======================

\vspace{2mm}
{\it Light CP-even Higgs boson}
\vspace{2mm}

In  the Minimal Supersymmetric Standard Model (MSSM)  there is correlation between the top quark mass and
the ratio of the vacuum expectation values, $tan\,\beta = v_2/v_1$\cite{GHKD}.  For the top mass in the range between 150 and
200 GeV  theoretically favored value of $tan\,\beta$ is $\sim$1.7 \cite{ CPW}. At  $M_t=$172 GeV (the central value)  the upper limit on the 
mass of the lightest CP-even Higgs boson $h$ is \cite{EFL}  
$$ M_h\lsim124\,GeV\,.$$
The current LEP data  restrict $ M_h$ from below  to the value 45 GeV.   A part of  this mass range  hits to the HERA
discovery range (\ref{DR}) and  therefore let us estimate the cross sections for production $h$ in
the  two-photon coherent $eA$ collisions.

To calculate the total cross section of $h$ production  it is necessary to integrate eq. (\ref{xseA})
over $W$ in the range
$$2 M_h<\,W\,<\sqrt{\frac{s_{ep}}{m_pR_A}}$$
with $ \sigma_{\gamma\gamma}$ from (\ref{gg}). Values of the partial decay widths $h\rightarrow\gamma\gamma$ 
can be calculated with the program HDECAY \cite{DKS}.
In  Fig.~4b presented the  cross sections  versus  the Higgs boson mass.  At a  realistic  integrated luminosity of the order
1/A pb$^{-1}$ and even two order higher the values of  the cross sections are far beyond the HERA sensitivity.

\vspace*{-3mm}
\section{Conclusions}
Analysis of the last section shows that in the two-photon coherent $eA$ collisions a number of the traditional processes
(with hadron and lepton production) can be studied at HERA  with high statistics. 
 However, only the restricted mass range    $M_{\tilde{l}}<13$ GeV can be explored  in processes with
SUSY particle production. Processes with production of MSSM Higgs particles will be 
unaccessible for study at the planed HERA luminosity.

The last chance to study  the Higgs bosons at HERA  provide \cite{kk} the general two doublet model (2HDM) \cite{GHKD},
since in framework of 2HDM light neutral Higgs particle is not rule out by present data.  
 Analysis of the light PC-even Higgs boson production in the two-photon $eA$ interactions will be performed elsewhere \cite{KL}.

\vspace{2mm} 
\noindent{\bf \large Acknowledgment.} The author is grateful to W. Krasny for stimulating discussions
and help.


\begin{thebibliography}{99}

\bibitem{BGMS}
V.M.~Budnev, I.F.~Ginzburg, G.V.~Meledin, V.G.~Serbo, Phys. Rep. {\bf C15} (1975) 181\zz
\vspace{-2mm}
\bibitem{Kessler1}
P. Kessler, Acta Phys. Austriaca, 41 (1975) 141\zz
\vspace{-2mm}
\bibitem{Field1}
J.H.~Field, in {\it Lecture Notes in Physics}:  Photon Photon Collisions, v. 191 (1983) 270\zz
\vspace{-2mm}
\bibitem{BB}
G.~Baur, C.A.~Bertulani, Phys. Rep. {\bf C163} (1988) 299\zz
\vspace{-2mm}
\bibitem{papag1}
E.~Papageorgiu, \NP {\bf A498} (1989) 593c; \Journal{\PR}{D40}{92}{1989}\zz
\vspace{-2mm}
\bibitem{Low}
F.~Low, Phys. Rev. {\bf 120} (1960) 582\zz
\vspace{-2mm}
\bibitem{Field2}
J.H.~Field, \Journal{\NP}{B168}{477}{1980}\zz
\vspace{-2mm}
\bibitem{ACK}
N.~Arteaga-Romero, C.~Carimalo, P.~Kessler, \Journal{\ZP}{C52}{289}{1991}\zz
\vspace{-2mm}
\bibitem{tau}
G.~Altarelli, QCD at Colliders, Preprint CERN-TH-95/196;\\
Proceedings of tau '94, \Journal{\NP}{B(Proc.Suppl.)}{1}{1995}\zz
\vspace{-2mm}
\bibitem{bussey}
P.J.~Bussey, B.~ Levtchenko, A.~Shumilin, in {\it Future Physics at HERA}, Proceedings of the Workshop,
DESY, Hamburg, May 1996\zz
\vspace{-2mm}
\bibitem{BKT}
S.J.~Brodsky, T. Kinoshita, H.~ Terazawa, \Journal{\PR}{D4}{1532}{1971}\zz
\vspace{-2mm}
\bibitem{GHKD}
J.F.~Gunion, H.E.~Haber, G.~Kane, S.~Dawson, The Higgs Hunter's Guide,
Addison-Wesley, Reading 1990\zz
\vspace{-2mm}
\bibitem{CPW}
M.~Carena, S.~Pokorski, C.E.M.~Wagnet, \Journal{\NP}{B406}{59}{1993}\zz
\vspace{-2mm}
\bibitem{EFL}
J.~Ellis, G.L.~Fogli, E.~Lisi, Preprint CERN-TH/95-202\zz
\vspace{-2mm}
\bibitem{DKS}
A.~Djouadi, J.~Kalinowski, M.~Spira, http://wwwcn.cern.ch/~mspira/prog \zz
\bibitem{kk}
\vspace{-2mm}
J.~Kalinowski, M.~Krawczyk, \Journal{\PL}{B361}{66}{1995}; Acta phys. Polon. 
{\bf B 27} (1996) 961; M. Krawczyk, talk at Moriond'96; 
D.~Choudhury, M.~Krawczyk,  Preprint MPI-PTh/96-46, IFT-96/13 ( hep-ph/9607271 )
\bibitem{KL}
\vspace{-2mm}
M.~Krawczyk, B.B.~Levtchenko, in preparation
\end{thebibliography}
\end{document}